\documentclass{PoS}

\usepackage{amsfonts,amsmath,amssymb,bm}
\usepackage{graphicx} 
\usepackage{pstricks}
\usepackage{subfig}

\newcommand{\be}{\begin{equation}}
\newcommand{\bea}{\begin{eqnarray}}
\newcommand{\ee}{\end{equation}}
\newcommand{\eea}{\end{eqnarray}}

\def\s#1{{\scriptscriptstyle #1}}
\def\n#1{({\it #1}\hspace{0.02cm})}

\def\noeq#1{(\ref{#1})}
\def\1eq#1{Eq.~(\ref{#1})}
\def\2eqs#1#2{Eqs.~(\ref{#1}) and~(\ref{#2})}
\def\3eqs#1#2#3{Eqs.~(\ref{#1}), (\ref{#2}) and~(\ref{#3})}
\def\4eqs#1#2#3#4{Eqs.~(\ref{#1}), (\ref{#2}), (\ref{#3}) and~(\ref{#4})}

\def\fig#1{Fig.~\ref{#1}}
\def\ie{{\it i.e.}, }
\def\eg{{\it e.g.}, }

\def\quark{\widehat{X}}
\def\qm{{\cal M}}
\def\pslash{p\hspace{-0.18cm}\slash}

\title{Unquenching the infrared sector of QCD}

\ShortTitle{Unquenching the infrared sector of QCD}

\author{\speaker{Daniele Binosi}\\
        European Centre for Theoretical Studies in Nuclear
Physics and Related Areas (ECT*) and Fondazione Bruno Kessler, \\Villa Tambosi, Strada delle
Tabarelle 286, 
I-38123 Villazzano (TN)  Italy\\
        E-mail: \email{binosi@ectsar.eu}}

\abstract{
We discuss the recent progress made in understanding how 
some fundamental results valid in the quenched infrared sector of non-Abelian Yang-Mills theories generalize to the unquenched case. In particular, we derive in the continuum an equation that allows to predict how the momentum dependence of the quenched gluon propagator is affected by the presence of a small number of dynamical quarks. In addition, we present the most recent lattice simulations of the (Landau gauge) gluon and ghost two-point sector obtained using  gauge field configurations  with two light and two light plus two heavy twisted-mass quark flavors. 
A comparison between the results obtained by these two complementary methods is then carried out.
}

\FullConference{Xth Quark Confinement and the Hadron Spectrum,\\
		October 8-12, 2012\\
		TUM Campus Garching, Munich, Germany}

\begin{document}

Our understanding of the infrared (IR) sector of non-Abelian Yang-Mills theories has considerably improved over the last few years. In particular, ab-initio {\it quenched} simulations~\cite{Cucchieri:2007md,Bogolubsky:2009dc} performed on large volumes lattices have unequivocally demonstrated that the (Landau gauge) gluon propagator and ghost dressing function saturate in the (deep) IR region.  Within the continuum formulation of the theory, these
lattice results are in agreement with, \eg the solutions of the
corresponding all-order Schwinger-Dyson equations
(SDEs)~\cite{Aguilar:2008xm,Boucaud:2008ky}  and exact
renormalization group  equations~\cite{Fischer:2008uz}. This has caused a paradigmatic shift among practitioners: the gluon is now thought to acquire a momentum-dependent mass $m(q^2)$ whose magnitude can be large at IR momenta, but rapidly vanishes with increasing spacelike momenta (\ie $q^2\gg\Lambda_\s{\rm QCD}^2$), thereby maintaining full accord with perturbative QCD. 

Here we report on our recent studies on the extension of these quenched results to the case in which dynamical quarks are present, both in the continuum as well as on the lattice. In particular:
\begin{itemize}

\item From the continuum side, we describe, within the general formalism based  on the  pinch technique
(PT)~\cite{Cornwall:1981zr,Cornwall:1989gv,Binosi:2002ft,Binosi:2003rr,Binosi:2009qm}  
and  the  background  field method  (BFM)~\cite{Abbott:1980hw} and known in the literature as the PT-BFM framework~\cite{Aguilar:2006gr,Binosi:2007pi,Binosi:2008qk}, how one can predict the effect that the presence of a small number of dynamical quarks will have on the shape of the gluon propagator, starting from the quenched results~\cite{Aguilar:2012rz};

\item From the lattice side, we will present the most recent data obtained in~\cite{Ayala:2012pb} where the full non-perturbative ghost and gluon two-point sector was computed by using gauge field configurations (gauge fixed in the Landau gauge) with two light and two light plus two heavy twisted-mass quark flavors.

\end{itemize}

\noindent A preliminary comparison between the results obtained will be finally presented.

\section{Schwinger-Dyson results}

In the PT-BFM framework it is possible to develop an approximate  method for computing nonperturbatively  the effects on the gluon propagator induced by the presence of a small number of dynamical quarks~\cite{Aguilar:2012rz}. 

To see how this can be done, let us start by observing that  when dynamical quarks are included in the calculation of a particular Green's function, there are two effects that naturally occur: \n{i} a new set of  diagrams  involving quark loops explicitly appears in the unquenched version of the corresponding SDE; \n{ii} the original quenched graphs get also modified, given that the presence of the quarks will in turn modify propagators, vertices and kernels that  appear inside them. In particular, specializing to the case of the (PT-BFM) SDE describing the gluon propagator,  one observes that in the full QCD case the original diagrams constituting the quenched version of the SDE, get supplemented with an extra graph containing the quark loop; however, the inclusion of this extra term will also affect indirectly the original (quenched) graphs since nested quark loops will now appear in them.

To proceed further, one then {\it assumes} that the quark-loop diagram captures the leading correction term, while additional corrections stemming  from diagrams with nested quark loops are  subleading. Neglecting these latter terms in front of the former, allows therefore to express the unquenched propagator as the sum of two contributions: the quenched propagator we started from plus the quark loop diagram. Then, in the PT-BFM framework the unquenched gluon propagator $\Delta_{\s Q}(q^2)$  can be obtained from the quenched propagator $\Delta(q^2)$ through the master formula (Euclidean space)~\cite{Aguilar:2012rz}
\be
\Delta_{\s{Q}}(q^2) = \frac{\Delta(q^2)}
{1 + \left\{ \quark(q^2) \left[1+G(q^2)\right]^{-2}+ \lambda^2(q^2)\right\}\Delta(q^2)}.
\label{mastformeuc}
\ee
Let us briefly recall what the different terms appearing on the right-hand side of the equation above correspond to.

To begin with,  $\widehat{X}(q^2)$ represents the PT-BFM scalar cofactor resulting from the calculation of the quark loop diagram; specifically, one has 
\be
\widehat{X}(q^2)=-\frac{g^2}{6}\!\int_k\mathrm{Tr}\left[
\gamma^\mu S(k)\widehat\Gamma_\mu(k+q,-k,-q)S(k+q)\right],
\label{qse}
\ee
where the $d$-dimensional integral (in dimensional regularization) is defined according to\linebreak \mbox{$\int_{k}\equiv\frac{\mu^{\epsilon}}{(2\pi)^{d}}\!\int\! \mathrm{d}^d k$},
with $d= 4-\epsilon$ and $\mu$ the 't Hooft mass. In addition, $S$ represents the full quark  propagator, defined as
\be
S^{-1}(p)=-i\left[A(p)\pslash-B(p)\right]=-iA(p)\left[\pslash-\qm(p)\right].
\label{qprop}
\ee
The ratio $\qm(p)=B(p)/A(p)$ represents the dynamical quark mass function; finally, the vertex $\widehat\Gamma_\nu$ is  the full PT-BFM quark-gluon vertex, which, due to its characteristic Abelian-like nature, satisfies the linear Ward identity (WI)
\be
ip_3^\mu\widehat\Gamma_\mu(p_1,p_2,p_3)=S^{-1}(p_1)-S^{-1}(-p_2).
\label{WI}
\ee
The appearance of  vertices satisfying linear WIs as opposed to the usual nonlinear Slavnov-Taylor identities (STIs) satisfied by conventional vertices, represents the distinctive feature of the PT-BFM framework, and simplifies the problem considerably, as it allows to employ for such vertices Abelian-like Ans\"atze. 

Next, the function $G(q^2)$ is a special Green's function particular to the PT-BFM which achieves the conversion from the PT-BFM gluon two-point sector to the conventional one~\cite{Binosi:2007pi,Binosi:2008qk}. In the Landau gauge $G(q^2)$ is known to coincide with the Kugo-Ojima function~\cite{Aguilar:2009pp,Grassi:2004yq}; in addition,  this function can be traded for the ghost dressing function $F$ thanks to the exact relation~\cite{Aguilar:2009pp,Grassi:2004yq}
$F^{-1}(q^2)=1+G(q^2)+L(q^2)$,
which, due to the fact that one has $L(q^2)\ll G(q^2)$ in the entire momentum range~\cite{Aguilar:2009pp},  lead to the approximate result $F^{-1}(q^2)\approx1+G(q^2)$.

Finally, the quantity $\lambda^2(q^2)$ is defined as
\be
\lambda^2(q^2)=m^2_\s{Q}(q^2)-m^2(q^2),
\label{lambda}
\ee
and corresponds to the difference between the dynamically generated gluon mass in the unquenched and the quenched cases, respectively. Evidently, the master equation~\noeq{mastformeuc} has been derived under the additional assumption that  the inclusion of light quark flavors (say, \eg an up and down type quarks,
with constituent masses of about \mbox{$300$ MeV})
will affect but not completely destabilize the mechanism responsible for 
the generation of a dynamical gluon mass, and that their effect may be considered as a ``perturbation'' 
to the quenched case (the inclusion of loops containing heavier quarks  has been shown to give rise to numerically suppressed contributions, consistent with the notion of decoupling~\cite{Aguilar:2012rz}).

One can next use  the quenched $SU(3)$ results obtained in large volume lattice simulations~\cite{Bogolubsky:2009dc} in order to evaluate the propagator $\Delta(q^2)$ and the ghost dressing function $F(q^2)$ [and therefore indirectly the special function $G(q^2)$ due to the relation between them].  Thus, what is left to be computed are the quantities $\widehat{X}(q^2)$ and $\lambda^2(q^2)$; one proceeds then as follows: 

\begin{itemize}

\item As far as the quark loop $\widehat{X}(q^2)$ is concerned, its evaluation comprises two steps. First, the  nonperturbative behavior of  the functions $A(p)$
and  ${\cal M}(p)$   appearing  in  the   definition  of  the   full  quark
propagator is obtained by solving  the quark gap equation. 
Particular care is needed for approximating  the non-Abelian  quark-gluon vertex $\Gamma_\mu$ appearing in this equation, since the latter is the conventional vertex (and {\it not} the PT-BFM $\widehat{\Gamma}_\mu$ vertex) , therefore satisfying the usual STI involving the ghost dressing function as well as the quark ghost scattering kernel~\cite{Marciano:1977su}. It turns out that the inclusion of this dependence in the vertex Ans\"atz used for solving the resulting nonlinear system  of  integral equations, is crucial in order to boost up the dynamically generated quark masses to phenomenologically acceptable values~\cite{Aguilar:2010cn}.
Once the functions $A(p)$ and ${\cal M}(p)$ are known, one can proceed to the calculation of the scalar factor $\widehat{X}(q^2)$, employing for the PT-BFM quark-gluon vertex  $\widehat{\Gamma}_\mu$ a suitable Abelian-like Ans\"atz, \eg the Ball-Chiu (BC) vertex~\cite{Ball:1980ay} or the Curtis-Pennington (CP)~\cite{Curtis:1990zs} vertex. Irrespectively from the vertex chosen, one has that  
in the $q\to0$ limit 
$\widehat{X}(0)=0$~\cite{Aguilar:2012rz}; thus,  the presence of fermions will not affect {\it directly} the saturation point of the (unquenched) gluon propagator (we will return on this important issue in the following point).

\item As for the quantity $\lambda^2(q^2)$ its first principle determination requires the knowledge of the equation which governs the dynamics of the effective gluon mass. This equation was derived in~\cite{Aguilar:2011ux,Binosi:2012sj} and turns out to be quite complex; in particular, it depends on the full gluon propagator: thus even though  $\widehat{X}(0)=0$, the  unquenched saturation point  will be different from the quenched one, since the overall shape of the propagator will be affected by the presence of dynamical quarks. The determination of the full $\lambda^2(q^2)$ using the aforementioned equation deserves a separate thorough study; in the ensuing analysis we will instead follow~\cite{Aguilar:2012rz},  restricting  ourselves to extracting  an approximate range for $\lambda^2$, by employing  a suitable extrapolation of the (unquenched) curves obtained from intermediate momenta towards the deep IR (thus we will effectively treating it as a constant).

\end{itemize}

\begin{figure}[!t]
\begin{minipage}[b]{0.45\linewidth}
\centering
\includegraphics[scale=0.4]{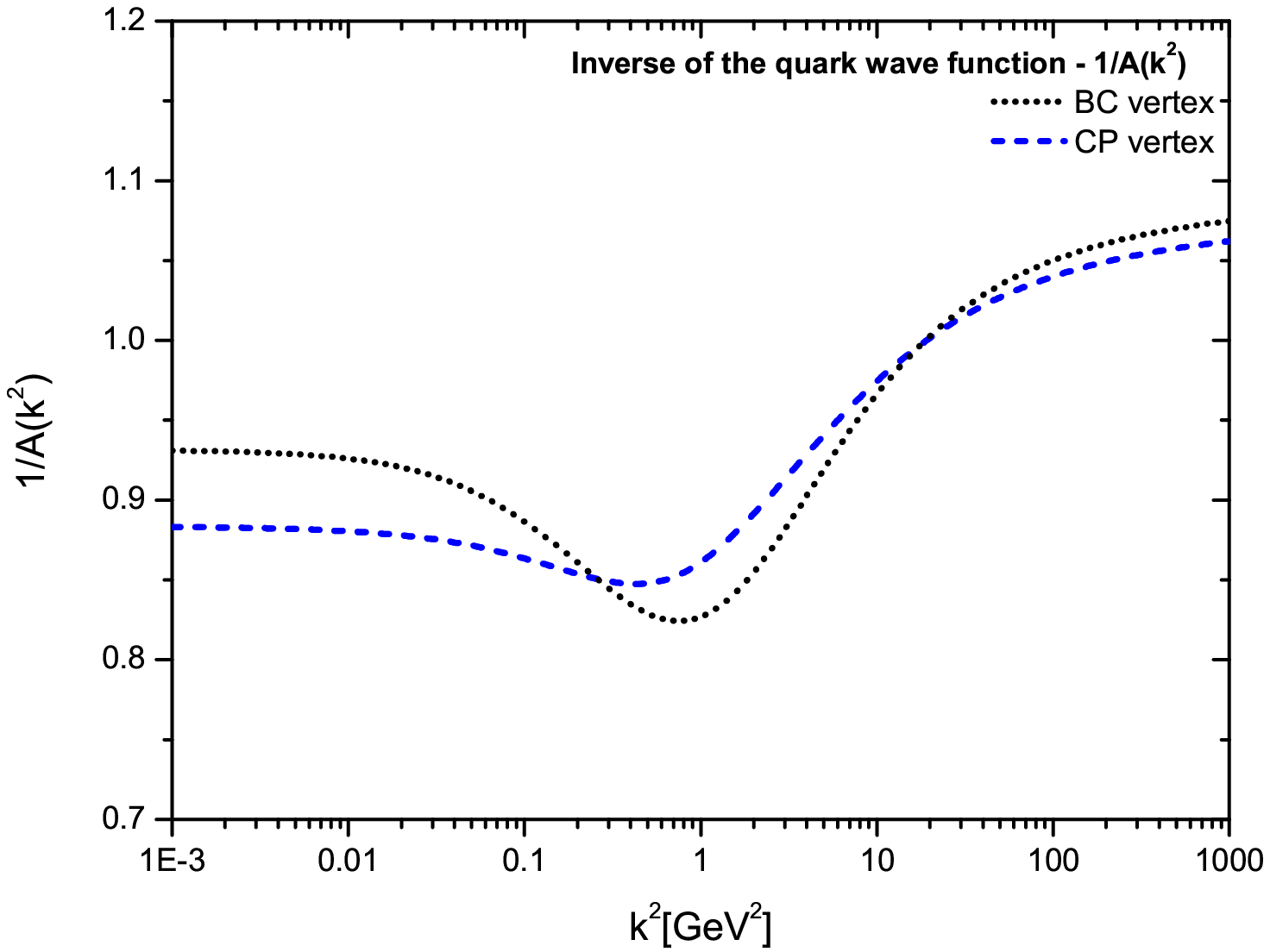}
\end{minipage}
\begin{minipage}[b]{0.50\linewidth}
\includegraphics[scale=0.41]{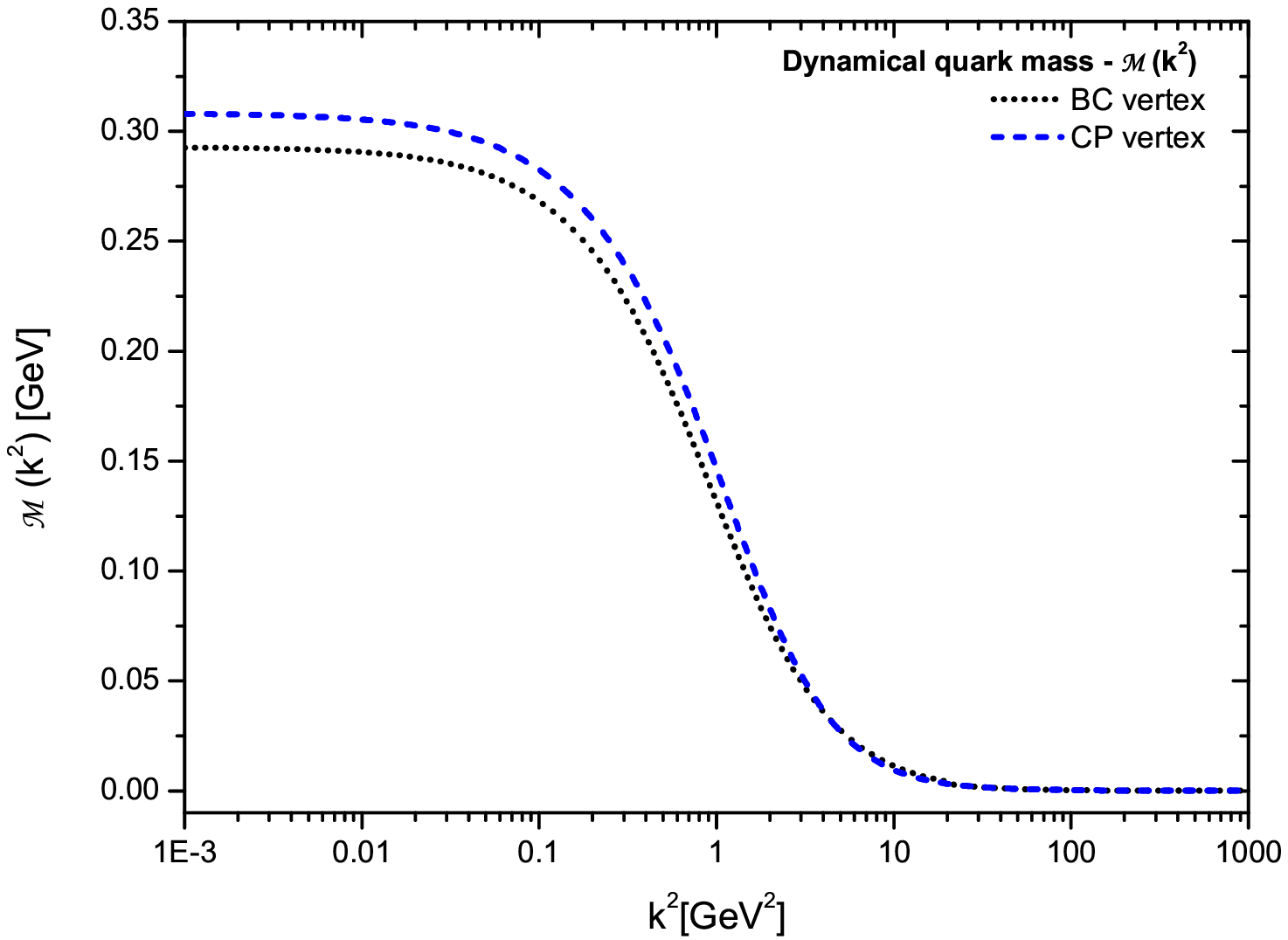}
\end{minipage}
\begin{minipage}[b]{0.45\linewidth}
\centering
\hspace{.18cm}
\includegraphics[scale=0.385]{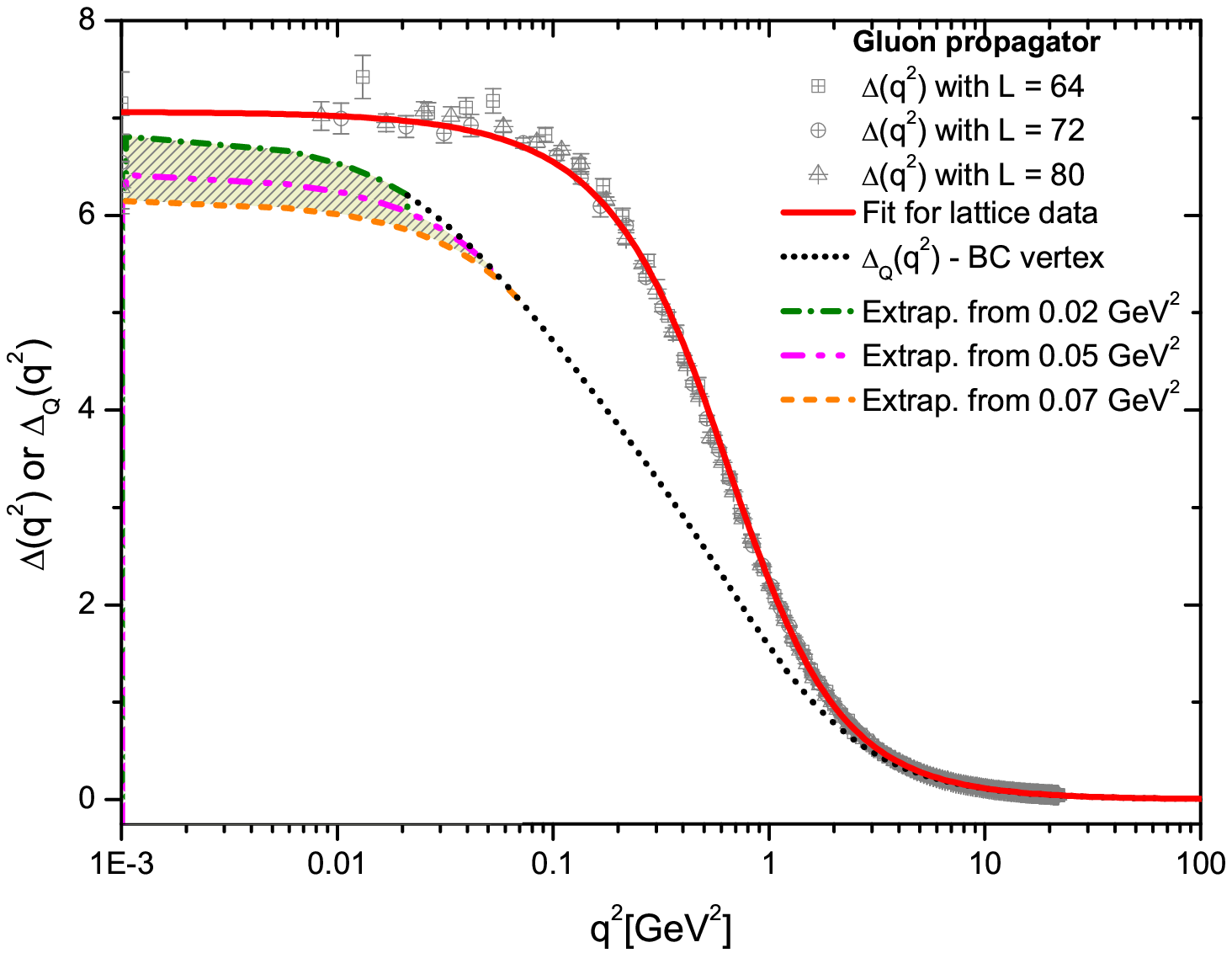}
\end{minipage}
\hspace{0.82cm}
\begin{minipage}[b]{0.50\linewidth}
\includegraphics[scale=0.39]{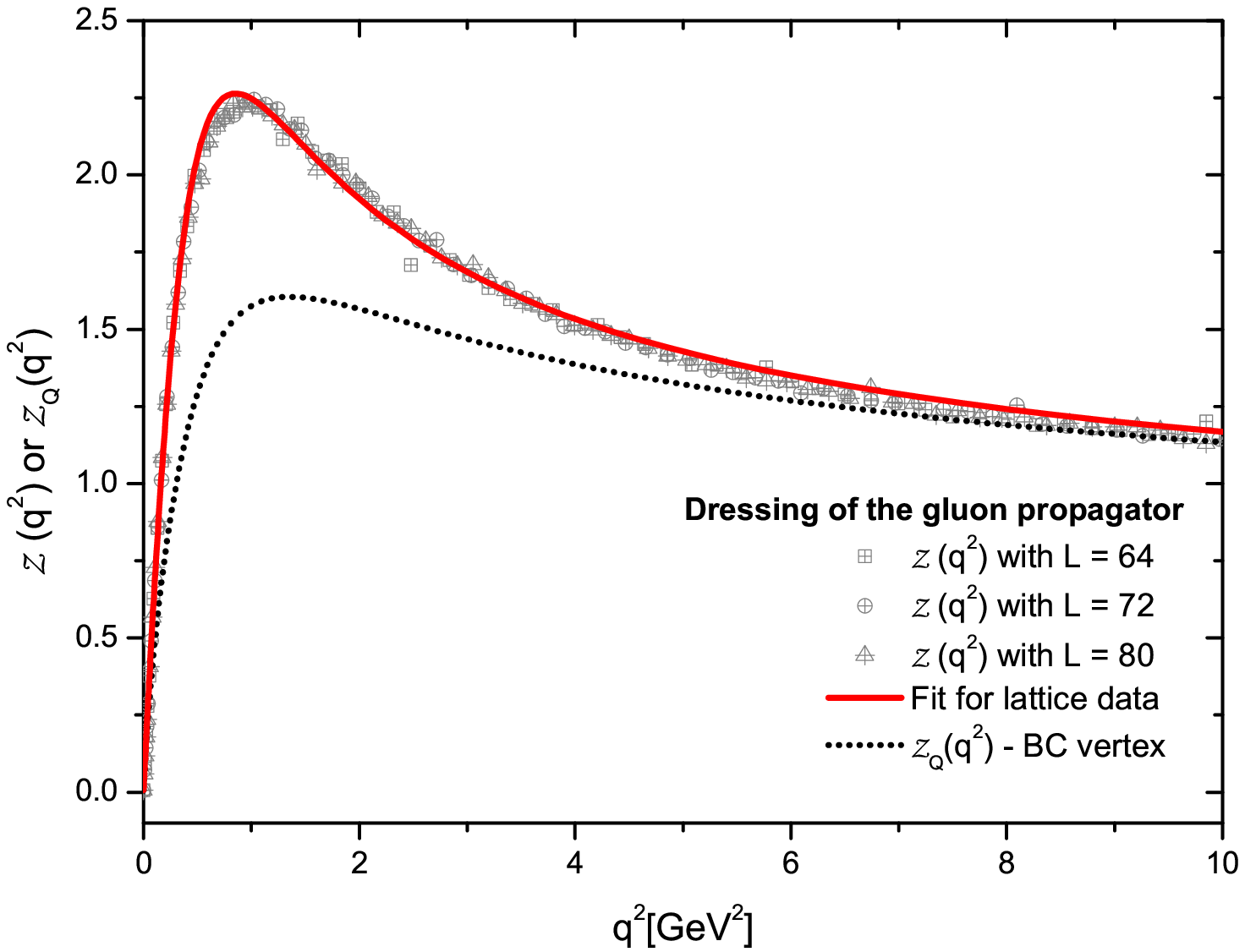}
\end{minipage}
\caption{\label{fig:unquenching1}({\it Top panels}) Solution of the quark gap equation:  $A^{-1}(p)$ (left) and dynamical quark mass \mbox{${\mathcal M}(p)$} (right) renormalized at $\mu= 4.3$ GeV; dotted black curves correspond to the BC vertex, while dashed blue curves to the  CP vertex. ({\it Bottom panels}) Comparison between the quenched and the unquenched gluon propagator (left) and dressing function (right), obtained using the BC vertex. The shaded striped band in the left plot shows
the possible values that $\Delta_{\s Q}(0)$ can assume at zero momentum depending on the extrapolation point used; in the case of the dressing function (which is basically insensible to the IR saturation point) we used a curve with an extrapolation point at $q^2=0.05\,\mbox{GeV}^2$. The quenched lattice results of~\cite{Bogolubsky:2009dc} are also displayed for comparison.}
\end{figure}

The results of the procedure so far described are shown in~ \fig{fig:unquenching1}, and can be summarized as follows.
The basic effect of the quark loops (two families with a constituent mass of the order of 300 MeV)  
is to suppress considerably the gluon propagator in the IR and intermediate momenta regions, while the ultraviolet tails increase, exactly as expected from the standard renormalization group analysis. In addition, the extrapolation procedure used (a cubic $B$-spline method applied starting from different momentum values) robustly predicts that the inclusion of light quarks makes the gluon propagator saturate at a lower point,  which can be translated into having a larger running gluon mass. As far as the gluon dressing function $q^2\Delta(q^2)$ is concerned, one observes a suppression of the intermediate momentum region peak, a feature already reported in early lattice studies.  

\section{Lattice results}

From the point of view of lattice simulations, a comprehensive quantitative study of the  gluon an ghost sectors in SU(3) QCD incorporating the effects stemming from the presence of two light and two light and two-heavy mass-twisted lattice flavors has been presented in~\cite{Ayala:2012pb} (see references therein for earlier lattice studies). The degenerate light quark masses used for the simulations ranged from 20 to 50 MeV, while the strange quark was roughly set to 95 MeV and the heavy charm to 1.51 GeV (in ${\overline{\rm MS}}$ at a scale of $2$ GeV). These values corresponds to a mass for the lightest pseudoscalar in the range approximately between 270 and 510 MeV;  finally, the biggest volume simulated corresponded to an asymmetrical box of roughly $3^3\times6$ [fm$^4$]. 

\begin{figure}[!t]
\begin{minipage}[b]{0.45\linewidth}
\includegraphics[scale=0.55]{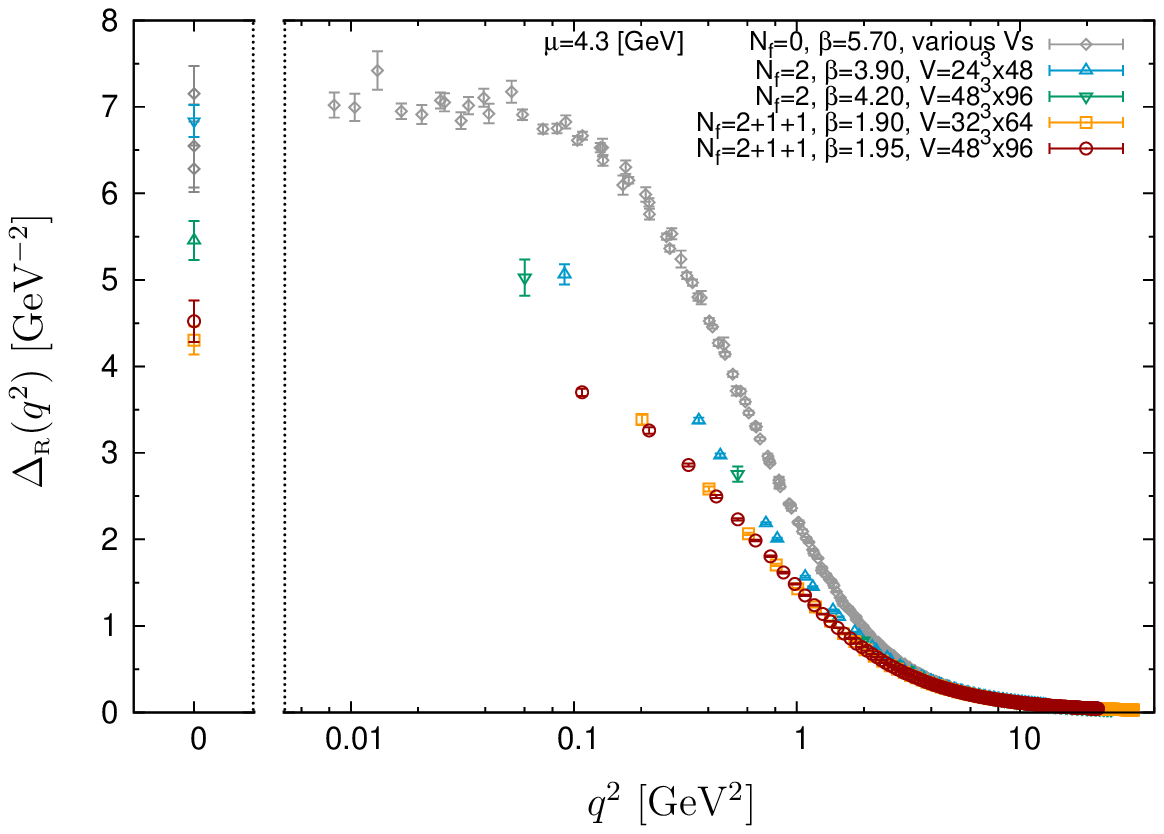}
\end{minipage}
\begin{minipage}[b]{0.45\linewidth}
\hspace{-1.2cm}
$\mbox{}$
\vspace{.15cm}
\includegraphics[scale=0.495]{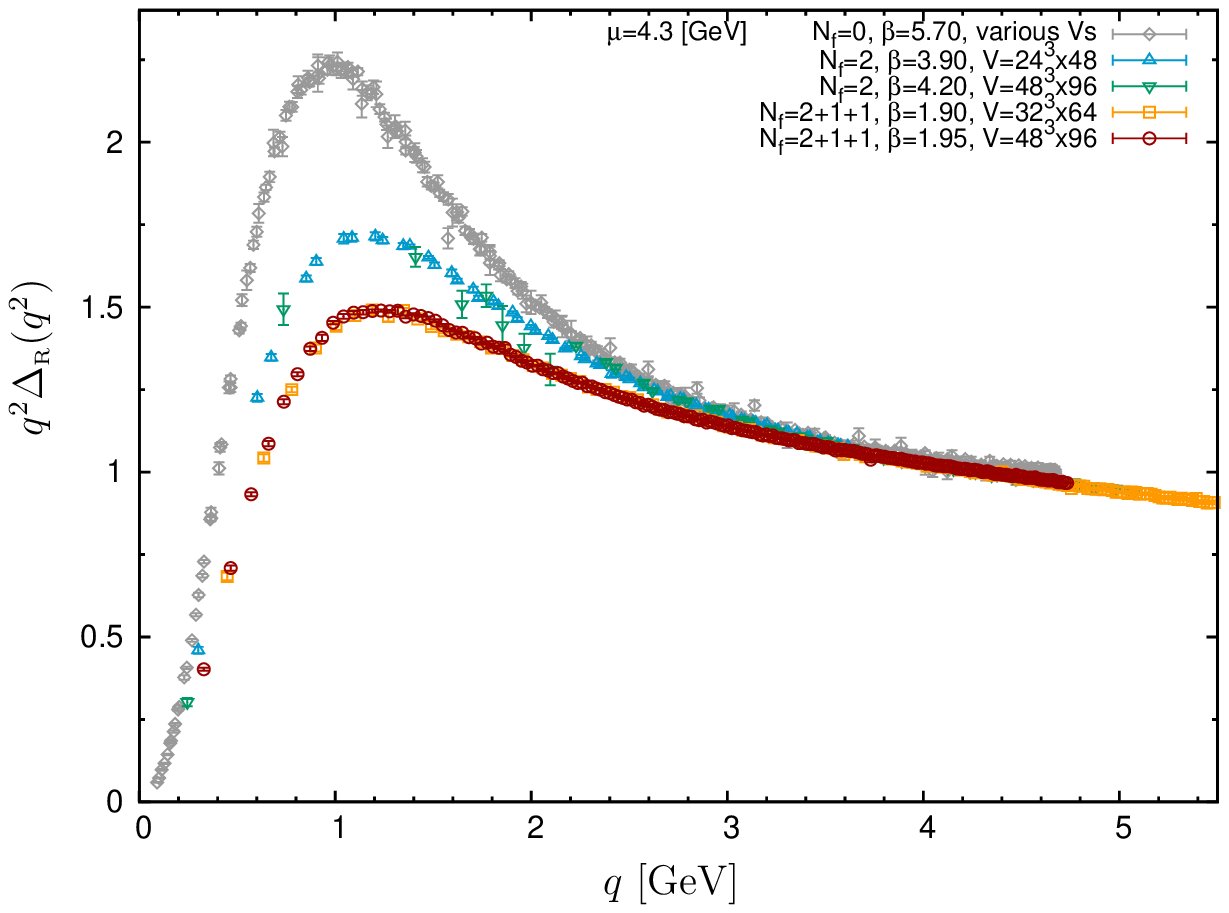}
\end{minipage}
\begin{minipage}[b]{0.45\linewidth}
$\mbox{}$\hspace{.75cm}
\includegraphics[scale=0.484]{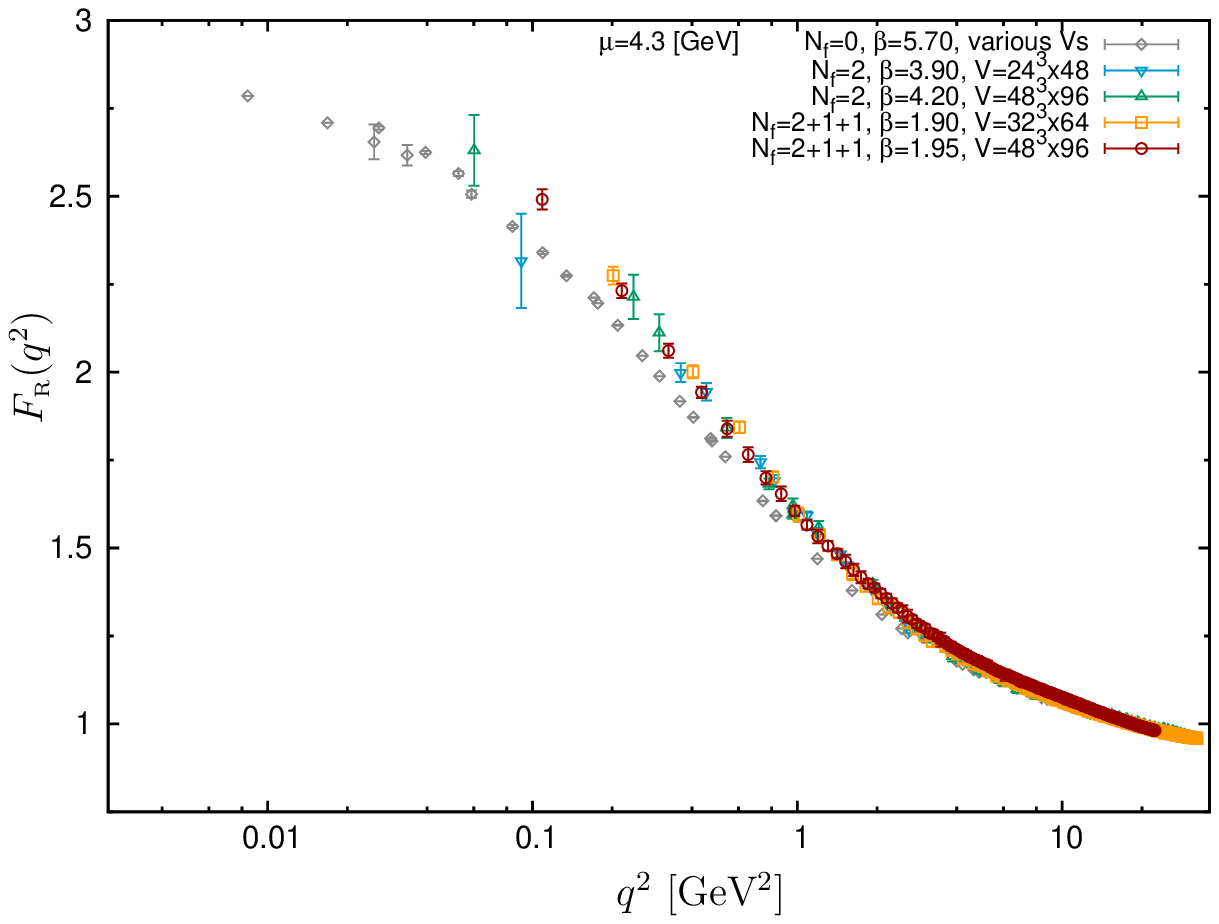}
\end{minipage}
\hspace{0.85cm}
\begin{minipage}[b]{0.50\linewidth}
\hspace{-0.70cm}
\includegraphics[scale=0.484]{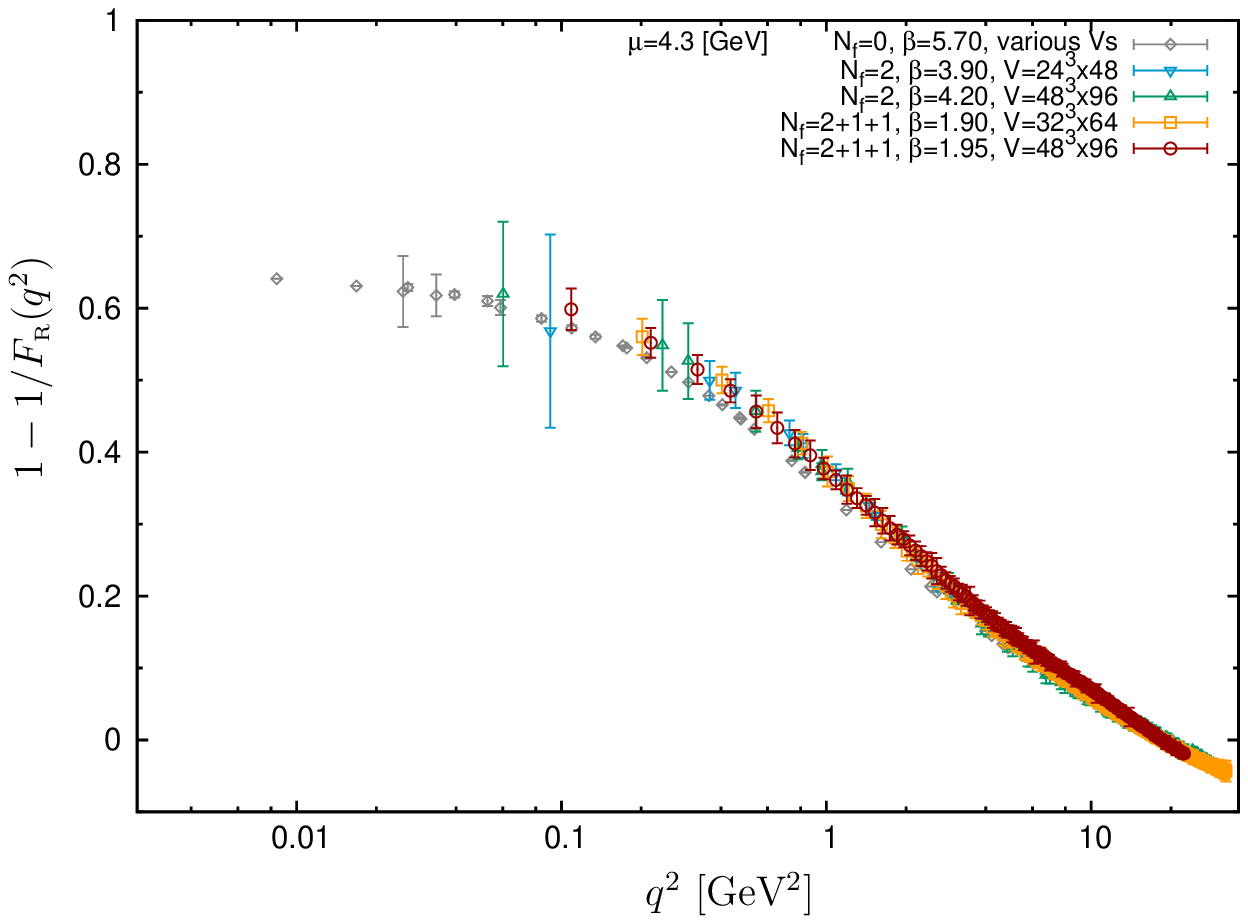}
\end{minipage}
\caption{\label{fig:unquenching2}({\it Top panels}) Landau gauge lattice results for the gluon propagator (left) and dressing function (right) for two light degenerate quarks (blue up triangles and green down triangles) and two light plus two heavy quarks (orange squares and red circles).  ({\it Bottom panels}) Lattice results for the ghost dressing $F(q^2)$ and $G(q^2)$ functions (color code as above). In all cases the data have been renormalized at $\mu=4.3$ GeV.}
\end{figure}

The results obtained (for an arbitrary
gauge copy selected through a gauge fixing algorithm using a
combination of over-relaxation and Fourier acceleration) are shown in~\fig{fig:unquenching2}, and are in full agreement with the SDE study discussed in the previous section. Indeed, one sees again that the effect of the presence of dynamical quarks on the gluon
propagator $\Delta$ is twofold: a suppression of both the ``swelling'' region at intermediate momenta and the saturation value in the deep IR. In addition, one observes that
the size of the effect increases as more light flavors are added.

On the other hand, the effect on the ghost dressing
function $F$ is much milder and is diametrically opposed to the one encountered for the gluon case, \ie it consists in a small
increase of the saturation point. Since in the SDE for the
ghost, dynamical quarks enters only indirectly, either through the gluon propagator or via higher loop corrections to the
gluon-ghost vertex, these results give support to the assumption (done in the SDE analysis of the previous section) that quark corrections due to nested quark loops can be expected to be small; in addition, it also shows that the approximation of the unquenched $F(q^2)$ and $G(q^2)$ with the corresponding quenched functions is valid to a very high degree of accuracy.

\section{Comparison and outlook}

\begin{figure}[!t]
\begin{minipage}[b]{0.45\linewidth}
\includegraphics[scale=0.55]{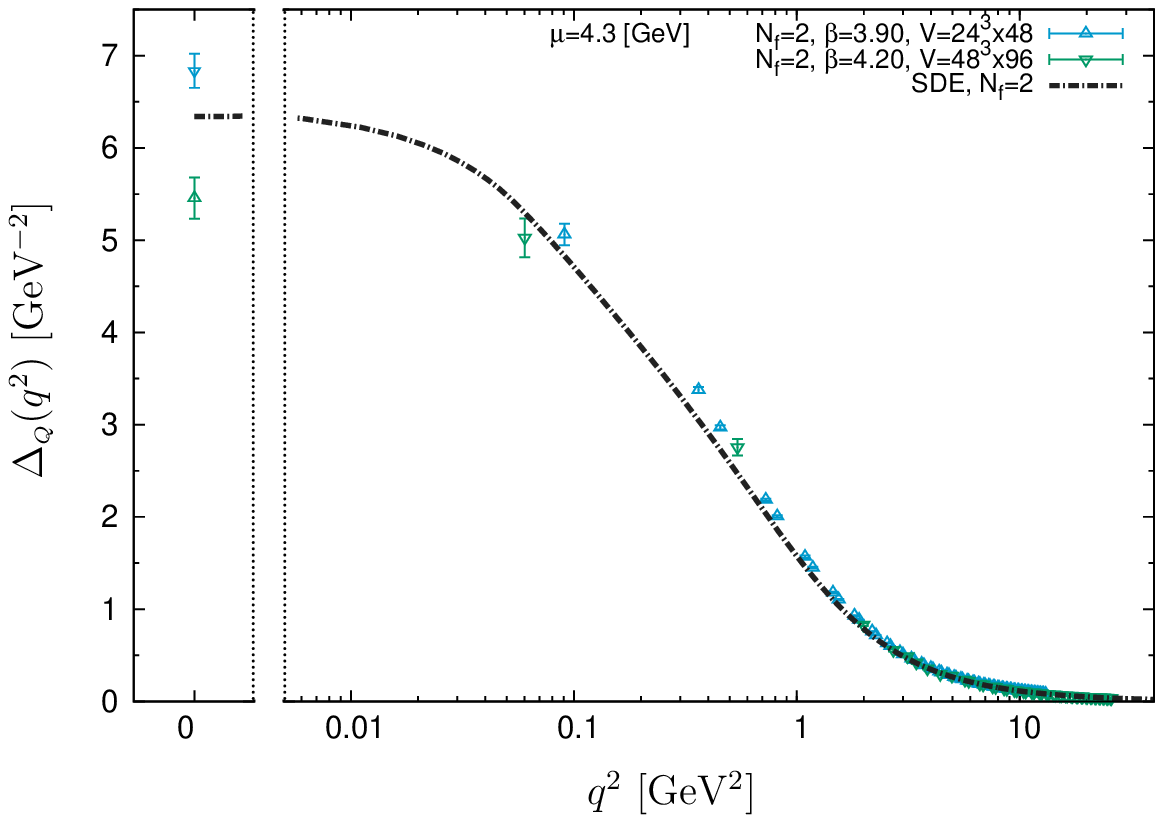}
\end{minipage}
\begin{minipage}[b]{0.45\linewidth}
\hspace{1cm}
\vspace{.25cm}
\includegraphics[scale=0.48]{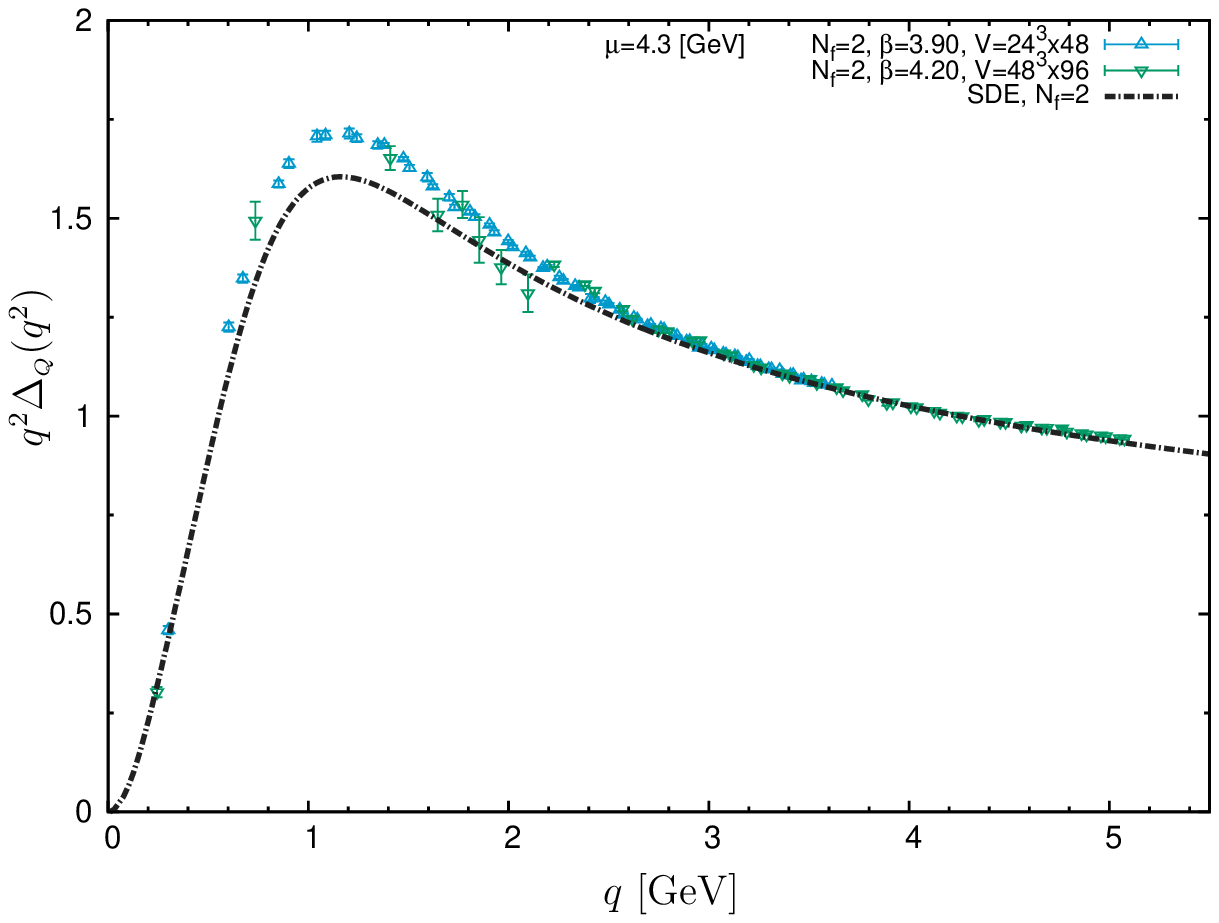}
\end{minipage}
\caption{\label{fig:comp} Comparison of  SDE  and lattice results for the case of two light fermions. The extrapolation of the SDE curve towards the IR is obtained starting from $q^2=0.05$ GeV$^2$.}
\end{figure}

In~\fig{fig:comp} we plot a comparison between the lattice results and the SDE computations for the case of two degenerate light quarks. The figure shows an excellent qualitative agreement  and a rather favorable quantitative agreement even at this level of approximation.

The discrepancies observed are due to the way in which $\lambda^2$ has been determined in~\cite{Aguilar:2012rz}. On the one hand, the extrapolation procedure overestimates the saturation point of the quenched propagator (notice that the blue lattice data point at zero momentum is affected by large volume artifacts, see~\cite{Ayala:2012pb}); on the other hand, the lack of knowledge of the full momentum dependence of $\lambda^2$ implies that the SDE curve (as is best seen in the plot of the dressing function) does not reproduce quantitatively the data in the intermediate momentum region (since the dynamical gluon mass is dropping pretty fast with $q^2$~\cite{Binosi:2012sj}, this is precisely the region in which we expect the biggest difference between the SDE and the lattice results). The determination of the full $\lambda^2(q^2)$ from the recently derived all-order mass equation~\cite{Binosi:2012sj} represents clearly the next step for this approach.

From the lattice point of view, on top of the obvious improvements which can be obtained by running simulations with bigger volumes and statistics  and smaller pseudoscalar masses,  it would be certainly very interesting to implement directly the simulations of PT-BFM  Green's functions using the canonical transformation technique recently developed in~\cite{Binosi:2012st}.


\begin{thebibliography}{10}

\bibitem{Cucchieri:2007md}
A.~Cucchieri and T.~Mendes, {\it {What's up with IR gluon and ghost propagators
  in Landau gauge? A puzzling answer from huge lattices}},
  \href{http://xxx.lanl.gov/abs/0710.0412}{{\tt 0710.0412}}.

\bibitem{Bogolubsky:2009dc}
I.~Bogolubsky, E.~Ilgenfritz, M.~Muller-Preussker, and A.~Sternbeck, {\it
  {Lattice gluodynamics computation of Landau gauge Green's functions in the
  deep infrared}},  {\em Phys.Lett.} {\bf B676} (2009) 69--73,
  [\href{http://xxx.lanl.gov/abs/0901.0736}{{\tt arXiv:0901.0736}}].

\bibitem{Aguilar:2008xm}
A.~Aguilar, D.~Binosi, and J.~Papavassiliou, {\it {Gluon and ghost propagators
  in the Landau gauge: Deriving lattice results from Schwinger-Dyson
  equations}},  {\em Phys.Rev.} {\bf D78} (2008) 025010,
  [\href{http://xxx.lanl.gov/abs/0802.1870}{{\tt arXiv:0802.1870}}].

\bibitem{Boucaud:2008ky}
P.~Boucaud et~al., {\it {On the IR behaviour of the Landau-gauge ghost
  propagator}},  {\em JHEP} {\bf 06} (2008) 099,
  [\href{http://xxx.lanl.gov/abs/0803.2161}{{\tt arXiv:0803.2161}}].

\bibitem{Fischer:2008uz}
C.~S. Fischer, A.~Maas, and J.~M. Pawlowski, {\it {On the infrared behavior of
  Landau gauge Yang-Mills theory}},  {\em Annals Phys.} {\bf 324} (2009)
  2408--2437, [\href{http://xxx.lanl.gov/abs/0810.1987}{{\tt
  arXiv:0810.1987}}].

\bibitem{Cornwall:1981zr}
J.~M. Cornwall, {\it {Dynamical Mass Generation in Continuum QCD}},  {\em Phys.
  Rev.} {\bf D26} (1982) 1453.

\bibitem{Cornwall:1989gv}
J.~M. Cornwall and J.~Papavassiliou, {\it {Gauge Invariant Three Gluon Vertex
  in QCD}},  {\em Phys. Rev.} {\bf D40} (1989) 3474.

\bibitem{Binosi:2002ft}
D.~Binosi and J.~Papavassiliou, {\it {The pinch technique to all orders}},
  {\em Phys. Rev.} {\bf D66} (2002) 111901(R),
  [\href{http://xxx.lanl.gov/abs/hep-ph/0208189}{{\tt hep-ph/0208189}}].

\bibitem{Binosi:2003rr}
D.~Binosi and J.~Papavassiliou, {\it {Pinch technique selfenergies and vertices
  to all orders in perturbation theory}},  {\em J.Phys.G} {\bf G30} (2004) 203,
  [\href{http://xxx.lanl.gov/abs/hep-ph/0301096}{{\tt hep-ph/0301096}}].

\bibitem{Binosi:2009qm}
D.~Binosi and J.~Papavassiliou, {\it {Pinch Technique: Theory and
  Applications}},  {\em Phys.Rept.} {\bf 479} (2009) 1--152,
  [\href{http://xxx.lanl.gov/abs/0909.2536}{{\tt arXiv:0909.2536}}].

\bibitem{Abbott:1980hw}
L.~F. Abbott, {\it {The Background Field Method Beyond One Loop}},  {\em Nucl.
  Phys.} {\bf B185} (1981) 189.

\bibitem{Aguilar:2006gr}
A.~C. Aguilar and J.~Papavassiliou, {\it {Gluon mass generation in the PT-BFM
  scheme}},  {\em JHEP} {\bf 12} (2006) 012,
  [\href{http://xxx.lanl.gov/abs/hep-ph/0610040}{{\tt hep-ph/0610040}}].

\bibitem{Binosi:2007pi}
D.~Binosi and J.~Papavassiliou, {\it {Gauge-invariant truncation scheme for the
  Schwinger-Dyson equations of QCD}},  {\em Phys.Rev.} {\bf D77} (2008) 061702,
  [\href{http://xxx.lanl.gov/abs/0712.2707}{{\tt arXiv:0712.2707}}].

\bibitem{Binosi:2008qk}
D.~Binosi and J.~Papavassiliou, {\it {New Schwinger-Dyson equations for
  non-Abelian gauge theories}},  {\em JHEP} {\bf 0811} (2008) 063,
  [\href{http://xxx.lanl.gov/abs/0805.3994}{{\tt arXiv:0805.3994}}].

\bibitem{Aguilar:2012rz}
A.~Aguilar, D.~Binosi, and J.~Papavassiliou, {\it {Unquenching the gluon
  propagator with Schwinger-Dyson equations}},  {\em Phys. Rev.} {\bf D86}
  (2012) 014032, [\href{http://xxx.lanl.gov/abs/1204.3868}{{\tt
  arXiv:1204.3868}}].

\bibitem{Ayala:2012pb}
A.~Ayala, A.~Bashir, D.~Binosi, M.~Cristoforetti, and J.~Rodriguez-Quintero,
  {\it {Quark flavour effects on gluon and ghost propagators}},  {\em
  Phys.Rev.} {\bf D86} (2012) 074512,
  [\href{http://xxx.lanl.gov/abs/1208.0795}{{\tt arXiv:1208.0795}}].

\bibitem{Aguilar:2009pp}
A.~Aguilar, D.~Binosi, and J.~Papavassiliou, {\it {Indirect determination of
  the Kugo-Ojima function from lattice data}},  {\em JHEP} {\bf 0911} (2009)
  066, [\href{http://xxx.lanl.gov/abs/0907.0153}{{\tt arXiv:0907.0153}}].

\bibitem{Grassi:2004yq}
P.~A. Grassi, T.~Hurth, and A.~Quadri, {\it {On the Landau background gauge
  fixing and the IR properties of YM Green functions}},  {\em Phys. Rev.} {\bf
  D70} (2004) 105014, [\href{http://xxx.lanl.gov/abs/hep-th/0405104}{{\tt
  hep-th/0405104}}].

\bibitem{Marciano:1977su}
W.~J. Marciano and H.~Pagels, {\it {Quantum Chromodynamics: A Review}},  {\em
  Phys. Rept.} {\bf 36} (1978) 137.

\bibitem{Aguilar:2010cn}
A.~Aguilar and J.~Papavassiliou, {\it {Chiral symmetry breaking with lattice
  propagators}},  {\em Phys.Rev.} {\bf D83} (2011) 014013,
  [\href{http://xxx.lanl.gov/abs/1010.5815}{{\tt arXiv:1010.5815}}].

\bibitem{Ball:1980ay}
J.~S. Ball and T.-W. Chiu, {\it {Analytic Properties of the Vertex Function in
  Gauge Theories. 1.}},  {\em Phys.Rev.} {\bf D22} (1980) 2542.

\bibitem{Curtis:1990zs}
D.~C. Curtis and M.~R. Pennington, {\it {Truncating the Schwinger-Dyson
  equations: How multiplicative renormalizability and the Ward identity
  restrict the three point vertex in QED}},  {\em Phys. Rev.} {\bf D42} (1990)
  4165--4169.

\bibitem{Aguilar:2011ux}
A.~Aguilar, D.~Binosi, and J.~Papavassiliou, {\it {The dynamical equation of
  the effective gluon mass}},  {\em Phys.Rev.} {\bf D84} (2011) 085026,
  [\href{http://xxx.lanl.gov/abs/1107.3968}{{\tt arXiv:1107.3968}}].

\bibitem{Binosi:2012sj}
D.~Binosi, D.~Ibanez, and J.~Papavassiliou, {\it {The all-order equation of the
  effective gluon mass}},  {\em Phys.Rev.} {\bf D86} (2012) 085033,
  [\href{http://xxx.lanl.gov/abs/1208.1451}{{\tt arXiv:1208.1451}}].

\bibitem{Binosi:2012st}
D.~Binosi and A.~Quadri, {\it {The Background Field Method as a Canonical
  Transformation}},  {\em Phys.Rev.} {\bf D85} (2012) 121702,
  [\href{http://xxx.lanl.gov/abs/1203.6637}{{\tt arXiv:1203.6637}}].

\end{thebibliography}
\providecommand{\href}[2]{#2}\begingroup\raggedright\endgroup

\end{document}